\documentclass[aps,prl%
,preprint%
,superscriptaddress%
]{revtex4-1}

\usepackage{graphicx}
\usepackage{dcolumn}
\usepackage{bm}
\usepackage{braket}
\usepackage{amssymb,amsmath}
\usepackage{xcolor}
\newcommand{\angstrom}{\mbox{\normalfont\AA}}
\def\Vhrulefill{\leavevmode\leaders\hrule height 0.7ex depth
\dimexpr0.4pt-0.7ex\hfill\kern0pt}
\begin{document}


\title{Estimated values of the kinetic energy for liquid $^3$He}


\author{V. Zampronio}
\email[]{viniciuz@ifi.unicamp.br}
\affiliation{Instituto de F\'{i}sica Gleb Wataghin,
University of Campinas - UNICAMP, 13083-859 Campinas - SP, Brazil}
\author{S. A. Vitiello}
\email[]{vitiello@ifi.unicamp.br}
\affiliation{Instituto de F\'{i}sica Gleb Wataghin,
University of Campinas - UNICAMP, 13083-859 Campinas - SP, Brazil}
\affiliation{CENAPAD-SP,
University of Campinas - UNICAMP, 13083-889 Campinas - SP, Brazil}


\date{\today}

\begin{abstract}
The kinetic energy is estimated for the ground-state of liquid $^3$He at
equilibrium density.  The obtained value for this quantity, $10.16\pm0.05$
K/atom at density $0.0163~\angstrom^{-3}$, is in agreement with most of
the experimental data found in the literature.  This result resolves a
long-standing controversy between experimental and theoretical values of
this quantity.  The variational path integral method, an``exact'' quantum
Monte Carlo method extended for fermionic systems, is applied in the
calculations.  The results obtained are subjected only to the restrictions
imposed by a chosen nodal structure without any further approximation,
even for quantities that do not commute with the Hamiltonian. The required
fixed-node approximation entails an implementation that allows a more
effective estimation of the quantities of interest.  Total and potential
energies together with the radial distribution function are also computed.
\end{abstract}

\pacs{02.70.Ss- ,67.30.E-}

\keywords{}

\maketitle

We investigated properties of normal condensed $^3$He at the equilibrium
density
and compared to experimental values. Neither experimental or theoretical
quantities of this system are easily obtained. Direct experimental
information about
single-particle dynamical properties such as the mean kinetic energy
$\langle E_K\rangle$
\cite{bry16,dim98,azu95,moo85,sok85}
of this strongly
interacting Fermion system can be obtained by deep inelastic neutron
scattering.  These are challenging experiments since the absorption
cross-section for thermal neutrons is about three orders of magnitude higher
than for inelastic scattering. On the other hand, theories using quantum
Monte Carlo many-body methods must avoid the Fermion sign problem that
so far has resisted an entirely  satisfactory answer. Most of the
experiments report kinetic energies in the range of 8 to 11~K/atom
\cite{bry16,dim98,azu95,moo85,sok85},
whereas theory predicted values between 12 and 13~K/atom
\cite{maz04,mor97,man83}.
This is a small, but a significant difference for an
``exact'' method.

In calculations made at zero temperature, we employed the variational path
integral (VPI) method
introduced by
Ceperley \cite{cep95}, who computed the total energy of $^4$He at
equilibrium density.
This is a well established method, also known as path-integral
ground-state (PIGS),  employed in the recent investigation of a
variety of bosonic systems, see for instance references
\onlinecite{ros17,abo18,ber16}.
We extended the
method to deal with
fermionic systems, in order to estimate properties of
liquid $^3$He.  In this  approach, a projection to the ground-state of the
system is made from a given initial state using ideas of path-integrals
over imaginary time \cite{cep95}.  The employed projector and how it
is used in the VPI method is reminiscent of how particles are treated
in a path-integral Monte Carlo calculation.  The necklace describing a
particle can be thought of as having
been cut and the coordinates at the extremities
are assumed to be those
of a trial function. This is what we refer to
as an open
path or polymer.
Configurations associated to monomers at the middle of
long enough polymers allow one to estimate any quantity, regardless
whether their expected values are associated to operators that do or do
not commute with the Hamiltonian.  ``Exact'' values are always
obtained without the need for any extrapolation.  However, since we are
dealing
with a fermionic system, the usual  fixed-node approximation needs to be
used. In our context, configurations
associated to the trial function at each end of the polymers
need to be considered.
Results obtained for all quantities of interest are only subjected to the
restrictions imposed by a chosen nodal structure.

Our main aim is the
investigation of properties of liquid $^3$He
associated with operators that do
not commute with the Hamiltonian.  We especially want to study the
kinetic energy of these systems, since there are controversies between
experimental and theoretical results that continue up until the present
\cite{bry16}.  We show that the VPI method gives estimates that are in
agreement with most of the experimental results.

Ground-state properties estimated by the VPI method are made by
applying the imaginary time
evolution operator,
$\rho(\beta)=\exp\left(-\beta{\cal H}\right)$,
with ${\cal H}$ being the system Hamiltonian,
in an initial state $\ket{\Psi_T}$ to project out the ground-state
$\ket{\phi_0}$.
The state
$\ket{\phi(\beta)}=\rho(\beta)\ket{\Psi_T}$ converges
exponentially to $\ket{\phi_0}$ as $\beta$ increases.

The matrix element
$\rho(R,R',\beta)=\bra{R}\rho(\beta)\ket{R'}$,
propagates
configuration $R$ to $R'$ in a ``time" $ \beta$ \cite{cep95},
where $R$ stands for all particle coordinates.
It is written as the
exponential of the action integrated over all paths.
The integration can  be made by factorizing $\rho(\beta)$
into the product of $M$ projectors $\rho(\tau)$, $\tau=\beta/M$, and
using the convolution property
\begin{eqnarray}\label{densitymatrix}
\rho(R,R',\beta) = && \displaystyle\int dR_1\ldots dR_{M-1}\rho(R,R_1,\tau) \\ \nonumber
 && \times
\rho(R_1,R_2,\tau)\ldots\rho(R_{M-1},R',\tau).
\end{eqnarray}
\noindent The intermediary configurations or beads $R_n$, $n = 1, \ldots, M-1$,
can be seen as the set of atomic coordinates at ``time" $t = n\tau$.
The beads stand for a sort of discretization of the path
from $R$ to $R'$ in a ``time" $\beta$. Therefore the integration of Eq.(\ref{densitymatrix})
converges to the integration over all paths
if $\tau$ is small enough.
In this case,
it is possible to employ the primitive approximation,
\begin{equation}\label{primitiverho}
\rho(R'',R''',\tau)\approx\rho_0(R'',R''',\tau)
e^{-\frac{\tau}{2}\left[V(R'')+V(R''')\right]},
\end{equation}
\noindent where $V(R)$ is the potential energy of
configuration $R$, and $\rho_0(R'',R''',\tau)$ is the
projector of non-interacting atoms,
$\rho_0(R'',R''',\tau)\propto\exp[-(R''-R''')^2/4\lambda\tau]$,
where $\lambda$ is $\hbar^2/2m$.
The primitive approximation is accurate to the second order in $\tau$.
We also implemented calculations with the Suzuki pair approximation \cite{cue05,ros09},
which is a fourth order in $\tau$ approximation,
\begin{equation}\label{szkrho}
\rho(R_k,R_{l},\tau)\approx\rho_0(R_{k},R_{l},\tau)
\displaystyle\prod_{i<j}e^{-U(r^{(k)}_{ij},r^{(l)}_{ij})},
\end{equation}
\noindent $r^{(\cdot)}_{ij}$ is the relative distance
between atoms $i$ and $j$ within configuration $R_{\cdot}$,
if $k$ is even
\begin{eqnarray}
U(r^{(k)},r^{(l)})=\frac{\tau}{3}
\left[2v(r^{(k)})+v(r^{(l)})\right]+ \\ \nonumber
\frac{\tau^3\lambda}{9}\left[\frac{\partial v}{\partial r}(r^{(k)})\right]^2,
\end{eqnarray}
\noindent and if $k$ is odd then
$U(r^{(k)},r^{(l)})=(\tau/3)[v(r^{(k)})+2v(r^{(l)})]$;
$v(r)$ is the inter-atomic potential.

By substituting Eq.(\ref{primitiverho})
or Eq.(\ref{szkrho}) into Eq.(\ref{densitymatrix})
we obtain a formula for $\rho(R,R',\beta)$.
Any error introduced by one of these approximations can, in general,
be made smaller than the statistical uncertainties of the Monte Carlo
method. The choice of Eq.(\ref{primitiverho})
or Eq.(\ref{szkrho})
did not affect our results.

In a system made of identical Fermions such as the one we are interested in,
the expression for $\rho(R,R',\beta)$ needs to be anti-symmetric under
the permutation of any pair of particles in the configuration $R$\cite{cep95,cep96}.
However, if $\Psi_T(R)$ is anti-symmetric, it is possible to
incorporate the minus sign rising from odd permutations in $\rho(R,R',\beta)$
into $\Psi_T(R)$ since
\begin{eqnarray}
\rho(R,R',\beta)\Psi_T(R')= \\ \nonumber
(-1)^{n_p}\rho(R,{\cal P}R',\beta)\Psi_T(R')= \\ \nonumber
\rho(R,{\cal P}R',\beta)\Psi_T({\cal P}R'),
\end{eqnarray}
\noindent where ${\cal P}$ changes the coordinates of
$n_p$ particles in a given configuration. And so, after integration in
$R'$, all permutation will have the same result (more details will be
given elsewhere).

Any property of the system in its ground-state can
be estimated in a straightforward manner.
If a given property is associated to
an operator ${\cal O}$,
its expected value can be written as
\begin{align}
O(\beta) & \propto  \bra{\phi(\beta)}{\cal O}\ket{\phi(\beta)} \\
\nonumber
& =
\bra{\Psi_T}\rho(\beta){\cal O}\rho(\beta)\ket{\Psi_T},
\end{align}
or as
\begin{align}
\label{expected}
O(\beta) = \int dR_1\ldots dR_{2M+1}P(R_1,\ldots,R_{2M+1}) O_{L}^{{\cal X}},
\end{align}
in terms of the probability
distribution function
$P$ of a given path
\begin{align}\label{probability}
P(R_1,\ldots,R_{2M+1}) &\propto \Psi_T(R_1)\rho(R_1,R_2,\tau)\ldots \\
\nonumber
& \times
\ldots\rho(R_{2M},R_{2M+1},\tau)\Psi_T(R_{2M+1}).
\end{align}
In Eq.~(\ref{expected}),
$O_L^{{\cal X}}(R_.)$ is the local value of the operator at a
given bead and
the index ${\cal X}$ labels different estimators
this method can allow us to use. If ${\cal O}$ commutes with
the Hamiltonian,
by using
its coordinate representation
it is possible to estimate its value for a given configuration $R$ at the
end of the path through
$O_L^{{\cal E}}(R) = {\cal O}\Psi_T(R)/\Psi_T(R)$.
This is the local value of $\cal O$ evaluated for
configurations at the end of the path associated to
$\Psi_T(R)$.

An estimate
of the ``exact" average value
of ${\cal O}$, even if it does not
commute with the system Hamiltonian, can be obtained through the
so called direct estimator given in the coordinate representation by
$O_L^{{\cal D}}(R_i,R_{i+1})
={\cal O}\rho(R_i,R_{i+1},\tau)/\rho(R_i,R_{i+1},\tau)$,
applied at the polymer middle.
For efficiency, the best approach is to consider the average
value $O_L^{{\cal D}}(R_i,R_{i+1})$ for $i=M$ and $i=M+1$.

For the total and kinetic energy estimates,
we can also use the
thermodynamic estimators $O^{\cal T}_L$
to consider configurations at the middle of the polymer.
In this context,
derivatives of
$\rho(\beta)$ with respect to $\beta$
and the mass $m$
are associated with the
total and kinetic energy respectively
\cite{cep95}.
For any of these estimators, care
must be taken when utilizing the Suzuki pair
approximation of Eq.(\ref{szkrho}), since the
operators must be inserted in odd beads \cite{ros09}. 

Since we want to investigate fermionic systems
the probability density
given by Eq.(\ref{probability})
can be negative. This is the sign problem common to most of the
ground-state Monte Carlo methods for fermionic systems.
Here we avoid this problem by rejecting
sampled paths where $\Psi_T(R_1)\Psi_T(R_{2M+1})<0$.
This is a fixed-node approximation that has
more degrees of freedom than the restriction $\Psi_T(R) > 0$
imposed when one applies an importance function transformation to
sample $\phi_0(R)\Psi_T(R)$, where $\phi_0(R)$ is unknown.
We believe that the extra degrees of freedom we have in this instance
improves
the exploration of the configuration space, especially to regions
where the nodal structure of $\Psi_T(R)$ is not identical
to that of the ground-state.

The system we consider is made of $N$ atoms of $^3$He inside a cubic box
with periodic boundary conditions applied to the faces of the box.
In our model,
the
atoms interact through the well-tested pairwise potential $v(r)$, HFD-B3-FCI1
\cite{azi95}, and
the Hamiltonian can be written as,
\begin{equation}\label{hamiltonian}
{\cal H} = \frac{1}{2m}\displaystyle\sum_{i=1}^N {\mathbf p}_i^2 + \displaystyle\sum_{i<j}^N v(r_{ij}),
\end{equation}
\noindent where $m$ is the $^3$He mass, ${\mathbf r}_i$ and ${\mathbf p}_i$
are respectively the coordinates and the momentum associated to a $i$-th atom
and $r_{ij}=|{\mathbf r}_i - {\mathbf r}_j|$.

It is interesting to experiment with different trial functions at the
end of the polymer. This allows us to investigate the convergence behavior
towards the exact ground-state of the quantities of interest. In this way,
two wave functions with different degrees of superposition with the
ground-state were considered. We performed
two series of independent runs, one for
each of the functions used at the extremities of the polymer.
The simplest function we have considered at the extremities
was the Jastrow-Slater (JS) wave function,
\begin{eqnarray}\label{JS}
\Psi_T(R) = e^{-\frac{1}{2}
\sum_{i<j} u(r_{ij})} \times \\ \nonumber 
{\rm det}_{\uparrow}\left(e^{i{\mathbf k}_l
\cdot
{\mathbf r}_m}\right)
{\rm det}_{\downarrow}\left(e^{i{\mathbf k}_l
\cdot
{\mathbf r}_n}\right),
\end{eqnarray}
where $u(r)=(b/r)^5$.
The nodal structure of this
wave function was improved by adding backflow correlations
in the Slater determinant\cite{sch81,pan89}.
These correlations are
introduced by a change in the particle coordinates,
${\mathbf r}_. \rightarrow {\mathbf r}_.
+ \sum_{j\ne \cdot} \eta(r_{\cdot j})
(\mathbf{r}_. - \mathbf{r}_j)$,
of the Slater determinant, where
\begin{equation}
\eta(r) = \lambda_B e^{-\left(\frac{r - s_B}{w_B}\right)^2}
+\frac{\lambda_B'}{r^3}.
\end{equation}
and $\lambda_B,\ s_B,\ w_B,\ \lambda_B'$ are parameters.
Three-body correlations\cite{sch81,pan89} were
also introduced
at the extremities of the open path.
Its functional form is
given by
\begin{equation}
\exp\left[-\frac{1}{2}\sum_{i<j} \tilde{u}(|\mathbf{r}_i-\mathbf{r}_j|)
-\frac{\lambda_T}{4}\sum_{l}\mathbf{G}(l)\cdot\mathbf{G}(l)\right],
\end{equation}
where
$\mathbf{G}(l)=\sum_{i\ne l}\xi(r_{ij})\mathbf{r}_{ij}$,
\begin{equation}
\xi(r) =
e^{-\left(\frac{r - s_T}{w_T}\right)^2}.
\end{equation}
and $s_T,\ w_T$ are parameters.
The pseudopotential
$\tilde{u}(r)=u(r)-\lambda_T\xi^2(r)r^2$
cancels
two-body factors arising from  $\mathbf{G}(l)$.
We refer to this improved wave function
as JS+BF+T.
In order for the wave function
to be periodic it is required that the
correlation functions
and its derivatives go smoothly to zero
at half of the side of the simulation box,
$L$. This can be achieved by the replacement
$f(r)\rightarrow f(r)+f(L-r)-2f(L/2)$, where
$f$ is either $u$, $\eta$ or $\xi$.

Our calculations were performed with $N=54$ atoms in a
non-polarized system
at the equilibrium density, $0.0163 \ \angstrom^{-3}$.
The sampling of the beads
were made by
the multi-Metropolis algorithm described in reference \cite{cep95}.
The configurations at the extremities of the path were sampled with the
usual Metropolis algorithm.

\begin{figure}
  \includegraphics{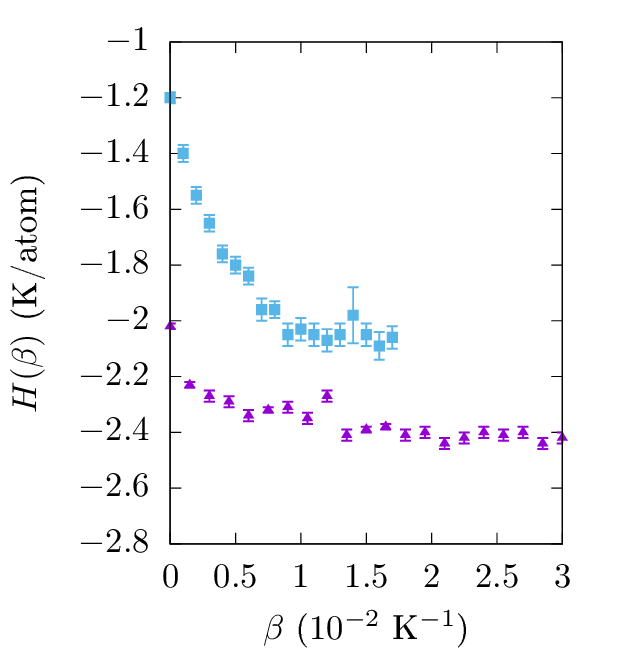}
  \caption{Total energy $H(\beta)$ calculated with the estimator at the end of the
  path. The points at $\beta = 0 \
  \text{K}^{-1}$ correspond to variational energies.  Squares represent
  the calculations with the JS wave function 
  and triangles stand for calculations using
  the JS+BF+T function.
  The parameters for this wave
  function are $b=2.99 \ \angstrom$, $\lambda_B=-0.14$, $\lambda_B'=-0.15$,
  $s_B=1.89 \ \angstrom$, $w_B=1.38 \
  \angstrom$, $\lambda_T=-1.8$,
   $s_T=1.69 \ \angstrom$ and $w_T=1.28 \ \angstrom$.}
  \label{energygraphic}
 \end{figure}

The total energy as a function of $\beta$, $H(\beta)$,
was calculated using the estimator at the end of the path
for the two different trial wave functions,
see Fig.~\ref{energygraphic}.
As $\beta$ increases the energy decreases, almost exponentially,
creating a sequence of
upper bound values to the ground-state energy. The results
show that improvements to the
trial wave function due to
backflow and three-body correlations
accelerate the convergence to the ground-state of the system.
The ground-state energy itself can only
be achieved if the nodal structure of
$\Psi_T(R)$ is identical to that of
the ground-state.
In this sense the
improvement in the nodal structure
of $\Psi_T(R)$ is noticeable due to the addition
of backflow correlations.

The tail ($\beta\ge 1.5\times10^{-2}$ K$^{-1}$)
of the curve in Fig.~\ref{energygraphic}
associated to the JS+BF+T wave function was fitted
to a constant straight line resulting in a total energy of $-2.41\pm 0.01$ K/atom,
which is a very good upper bound to the experimental data
$-2.47\pm 0.01$ K/atom \cite{rob64}.
From now on, all results we report 
are in reference to the results obtained from the wave function
above.

\begin{figure}
  \includegraphics{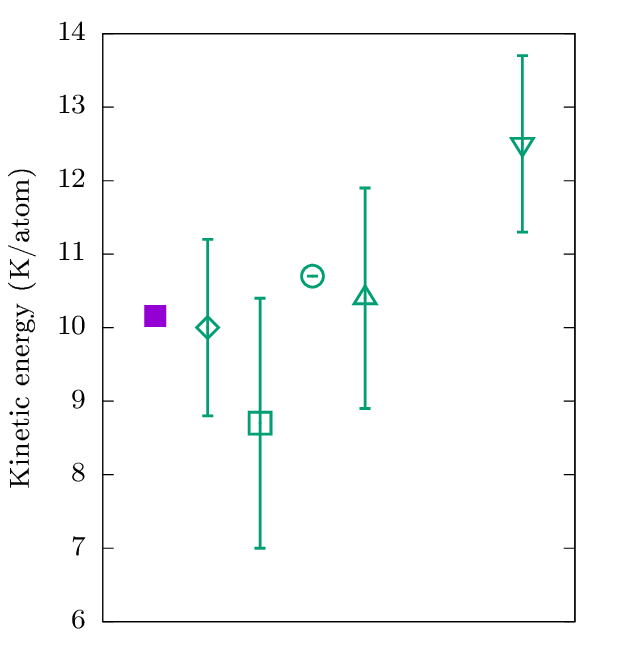}
  \caption{Comparison of our results for the kinetic energy
of $^3$He at the equilibrium density
obtained with projections of configurations from the JS+BF+T wave function, and
experimental data from the literature.
The symbols were horizontally displaced for the sake of clarity.
The full square stands for
our calculation; this symbol size is larger than the statistical
uncertainty.
The experimental data are plotted with
empty symbols:
square \cite{sok85}, circle \cite{moo85},
up triangle \cite{azu95} down triangle \cite{bry16},
and rhombus \cite{dim98}.
 \label{energygraphic2}}
 \end{figure}

For the estimation of the kinetic energy we use the
same procedure of considering all the converged values we have obtained for this
quantity. The straight line fit to these results gave us the value
we adopt for the ground-state kinetic energy,
$10.16\pm0.05$ K/atom.
In Fig.~\ref{energygraphic2}, we plotted this value
together with experimental data
from the literature for liquid $^3$He at equilibrium density.
Most of the experimental data lies in a range
from $8$ to $11$ K/atom, which is in excellent
agreement with our estimates, thus resolving
a long-standing disagreement between experimental data
and theoretical Monte Carlo calculations
For completeness
in Table \ref{energytable}, we give values of the
potential and kinetic energies
calculated with the direct and thermodynamic
estimators.

 \begin{table}
 \caption{Kinetic and potential energies in units of
 K/atom evaluated with the direct $(\cal D)$ and thermodynamic
 $(\cal T)$ estimators using configurations projected from
 the JS+BF+T wave function.\label{energytable}}
 \begin{ruledtabular}
 \begin{tabular}{lll}
\noalign{\smallskip}
Estimator & Kinetic Energy & Potential Energy \\
\noalign{\smallskip}\hline\noalign{\smallskip}
 ${\cal D}$ & $10.19\pm0.07$ & $-12.75\pm0.01$ \\
 ${\cal T}$ & $10.14\pm0.07$ & $-12.75\pm0.01$ \\  
 \end{tabular}
 \end{ruledtabular}
 \end{table}

\begin{figure}
  \includegraphics{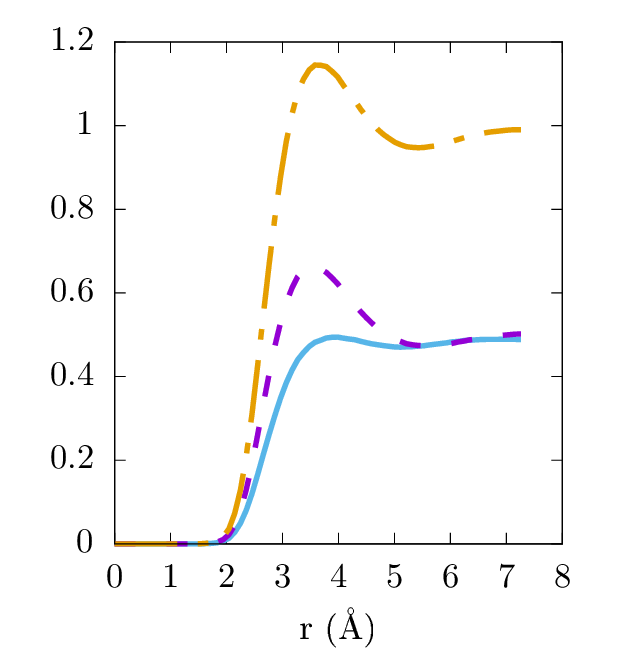}
  \caption{Radial distribution function of liquid $^3$He at equilibrium
  density.  The total radial distribution function is represented by the
  dashed-and-dotted line; the spin-resolved distribution functions for
  spin parallel and spin anti-parallel atoms are represented by the solid
  and dashed lines respectively.
\label{gr}}
\end{figure}

We have also calculated
the radial distribution of atoms
and its spin-resolved components for atoms with
parallel and anti-parallel spins.
These quantities were calculated with the direct estimator,
see Fig.~\ref{gr}.
The anti-parallel spin curve has a more pronounced  peak
because atoms with different spins do not suffer the Pauli
exclusion.

In summary,
as we have shown, the VPI approach to study the
ground-state of fermionic systems is robust and reliable.
Any quantity, associated with
operators that do or do not commute with the Hamiltonian
is readily estimated
without the need of extrapolations.
By avoiding them,
estimates can be obtained
completely free
from any possible bias introduced by
variational calculations.
Moreover,
a long-standing disagreement between experimental data
and theoretical calculations of the ground-state kinetic energy
was resolved.
To what degree the findings in this work will be reflected in other
systems is still an open question. Nevertheless,
this question is very important, because
many results for physical properties of great interest
were obtained in the literature using extrapolations.
\emph{Acknowledgments:}
The authors acknowledge financial support from the Brazilian
agencies Funda\c c\~ao de Amparo \`a Pesquisa do Estado de S\~ao Paulo
(FAPESP) and Conselho Nacional de Desenvolvimento Cient\'\i fico e
Tecnol\'ogico (CNPq). Part of the computations were performed at the
Centro Nacional de Processamento de Alto Desempenho em S\~ao Paulo
(CENAPAD-SP).


\begin{thebibliography}{19}%
\makeatletter
\providecommand \@ifxundefined [1]{%
 \@ifx{#1\undefined}
}%
\providecommand \@ifnum [1]{%
 \ifnum #1\expandafter \@firstoftwo
 \else \expandafter \@secondoftwo
 \fi
}%
\providecommand \@ifx [1]{%
 \ifx #1\expandafter \@firstoftwo
 \else \expandafter \@secondoftwo
 \fi
}%
\providecommand \natexlab [1]{#1}%
\providecommand \enquote  [1]{``#1''}%
\providecommand \bibnamefont  [1]{#1}%
\providecommand \bibfnamefont [1]{#1}%
\providecommand \citenamefont [1]{#1}%
\providecommand \href@noop [0]{\@secondoftwo}%
\providecommand \href [0]{\begingroup \@sanitize@url \@href}%
\providecommand \@href[1]{\@@startlink{#1}\@@href}%
\providecommand \@@href[1]{\endgroup#1\@@endlink}%
\providecommand \@sanitize@url [0]{\catcode `\\12\catcode `\$12\catcode
  `\&12\catcode `\#12\catcode `\^12\catcode `\_12\catcode `\%12\relax}%
\providecommand \@@startlink[1]{}%
\providecommand \@@endlink[0]{}%
\providecommand \url  [0]{\begingroup\@sanitize@url \@url }%
\providecommand \@url [1]{\endgroup\@href {#1}{\urlprefix }}%
\providecommand \urlprefix  [0]{URL }%
\providecommand \Eprint [0]{\href }%
\providecommand \doibase [0]{http://dx.doi.org/}%
\providecommand \selectlanguage [0]{\@gobble}%
\providecommand \bibinfo  [0]{\@secondoftwo}%
\providecommand \bibfield  [0]{\@secondoftwo}%
\providecommand \translation [1]{[#1]}%
\providecommand \BibitemOpen [0]{}%
\providecommand \bibitemStop [0]{}%
\providecommand \bibitemNoStop [0]{.\EOS\space}%
\providecommand \EOS [0]{\spacefactor3000\relax}%
\providecommand \BibitemShut  [1]{\csname bibitem#1\endcsname}%
\let\auto@bib@innerbib\@empty
\bibitem [{\citenamefont {Bryan}\ \emph {et~al.}(2016)\citenamefont {Bryan},
  \citenamefont {Prisk}, \citenamefont {Azuah}, \citenamefont {Stirling},\ and\
  \citenamefont {Sokol}}]{bry16}%
  \BibitemOpen
  \bibfield  {author} {\bibinfo {author} {\bibfnamefont {M.~S.}\ \bibnamefont
  {Bryan}}, \bibinfo {author} {\bibfnamefont {T.~R.}\ \bibnamefont {Prisk}},
  \bibinfo {author} {\bibfnamefont {R.~T.}\ \bibnamefont {Azuah}}, \bibinfo
  {author} {\bibfnamefont {W.~G.}\ \bibnamefont {Stirling}}, \ and\ \bibinfo
  {author} {\bibfnamefont {P.~E.}\ \bibnamefont {Sokol}},\ }\href@noop {}
  {\bibfield  {journal} {\bibinfo  {journal} {EPL}\ }\textbf {\bibinfo {volume}
  {115}},\ \bibinfo {pages} {6601} (\bibinfo {year} {2016})}\BibitemShut
  {NoStop}%
\bibitem [{\citenamefont {Dimeo}\ \emph {et~al.}(1998)\citenamefont {Dimeo},
  \citenamefont {Sokol}, \citenamefont {Azuah}, \citenamefont {Bennington},
  \citenamefont {Stirling},\ and\ \citenamefont {Guckelsberger}}]{dim98}%
  \BibitemOpen
  \bibfield  {author} {\bibinfo {author} {\bibfnamefont {R.~M.}\ \bibnamefont
  {Dimeo}}, \bibinfo {author} {\bibfnamefont {P.~E.}\ \bibnamefont {Sokol}},
  \bibinfo {author} {\bibfnamefont {R.~T.}\ \bibnamefont {Azuah}}, \bibinfo
  {author} {\bibfnamefont {S.~M.}\ \bibnamefont {Bennington}}, \bibinfo
  {author} {\bibfnamefont {W.~G.}\ \bibnamefont {Stirling}}, \ and\ \bibinfo
  {author} {\bibfnamefont {K.}~\bibnamefont {Guckelsberger}},\ }\href@noop {}
  {\bibfield  {journal} {\bibinfo  {journal} {Physica B}\ }\textbf {\bibinfo
  {volume} {241-243}},\ \bibinfo {pages} {952} (\bibinfo {year}
  {1998})}\BibitemShut {NoStop}%
\bibitem [{\citenamefont {Azuah}\ \emph {et~al.}(1995)\citenamefont {Azuah},
  \citenamefont {Stirling}, \citenamefont {Guckelsberger}, \citenamefont
  {Scherm}, \citenamefont {Bennington}, \citenamefont {Yates}, ,\ and\
  \citenamefont {Taylor}}]{azu95}%
  \BibitemOpen
  \bibfield  {author} {\bibinfo {author} {\bibfnamefont {R.~T.}\ \bibnamefont
  {Azuah}}, \bibinfo {author} {\bibfnamefont {W.~G.}\ \bibnamefont {Stirling}},
  \bibinfo {author} {\bibfnamefont {K.}~\bibnamefont {Guckelsberger}}, \bibinfo
  {author} {\bibfnamefont {R.}~\bibnamefont {Scherm}}, \bibinfo {author}
  {\bibfnamefont {S.~M.}\ \bibnamefont {Bennington}}, \bibinfo {author}
  {\bibfnamefont {M.~L.}\ \bibnamefont {Yates}}, , \ and\ \bibinfo {author}
  {\bibfnamefont {A.~D.}\ \bibnamefont {Taylor}},\ }\href@noop {} {\bibfield
  {journal} {\bibinfo  {journal} {J. Low Temp. Phys.}\ }\textbf {\bibinfo
  {volume} {101}},\ \bibinfo {pages} {951} (\bibinfo {year}
  {1995})}\BibitemShut {NoStop}%
\bibitem [{\citenamefont {Mook}(1985)}]{moo85}%
  \BibitemOpen
  \bibfield  {author} {\bibinfo {author} {\bibfnamefont {H.~A.}\ \bibnamefont
  {Mook}},\ }\href@noop {} {\bibfield  {journal} {\bibinfo  {journal} {Phys.
  Rev. Lett.}\ }\textbf {\bibinfo {volume} {22}},\ \bibinfo {pages} {2452}
  (\bibinfo {year} {1985})}\BibitemShut {NoStop}%
\bibitem [{\citenamefont {Sokol}\ \emph {et~al.}(1985)\citenamefont {Sokol},
  \citenamefont {Sköld}, \citenamefont {Price},\ and\ \citenamefont
  {Kleb}}]{sok85}%
  \BibitemOpen
  \bibfield  {author} {\bibinfo {author} {\bibfnamefont {P.~E.}\ \bibnamefont
  {Sokol}}, \bibinfo {author} {\bibfnamefont {K.}~\bibnamefont {Sköld}},
  \bibinfo {author} {\bibfnamefont {D.~L.}\ \bibnamefont {Price}}, \ and\
  \bibinfo {author} {\bibfnamefont {R.}~\bibnamefont {Kleb}},\ }\href@noop {}
  {\bibfield  {journal} {\bibinfo  {journal} {Phys. Rev. Lett.}\ }\textbf
  {\bibinfo {volume} {54}},\ \bibinfo {pages} {909} (\bibinfo {year}
  {1985})}\BibitemShut {NoStop}%
\bibitem [{\citenamefont {Mazzanti}\ \emph {et~al.}(2004)\citenamefont
  {Mazzanti}, \citenamefont {Polls}, \citenamefont {Boronat},\ and\
  \citenamefont {Casulleras}}]{maz04}%
  \BibitemOpen
  \bibfield  {author} {\bibinfo {author} {\bibfnamefont {F.}~\bibnamefont
  {Mazzanti}}, \bibinfo {author} {\bibfnamefont {A.}~\bibnamefont {Polls}},
  \bibinfo {author} {\bibfnamefont {J.}~\bibnamefont {Boronat}}, \ and\
  \bibinfo {author} {\bibfnamefont {J.}~\bibnamefont {Casulleras}},\
  }\href@noop {} {\bibfield  {journal} {\bibinfo  {journal} {Phys. Rev. Lett.}\
  }\textbf {\bibinfo {volume} {92}},\ \bibinfo {pages} {085301} (\bibinfo
  {year} {2004})}\BibitemShut {NoStop}%
\bibitem [{\citenamefont {Moroni}\ \emph {et~al.}(1997)\citenamefont {Moroni},
  \citenamefont {Senatore},\ and\ \citenamefont {Fantoni}}]{mor97}%
  \BibitemOpen
  \bibfield  {author} {\bibinfo {author} {\bibfnamefont {S.}~\bibnamefont
  {Moroni}}, \bibinfo {author} {\bibfnamefont {G.}~\bibnamefont {Senatore}}, \
  and\ \bibinfo {author} {\bibfnamefont {S.}~\bibnamefont {Fantoni}},\ }\href
  {\doibase https://doi.org/10.1103/PhysRevB.55.1040} {\bibfield  {journal}
  {\bibinfo  {journal} {Phys. Rev. B}\ }\textbf {\bibinfo {volume} {55}},\
  \bibinfo {pages} {1040} (\bibinfo {year} {1997})}\BibitemShut {NoStop}%
\bibitem [{\citenamefont {Manousakis}\ \emph {et~al.}(1983)\citenamefont
  {Manousakis}, \citenamefont {Fantoni}, \citenamefont {Pandharipande},\ and\
  \citenamefont {Usmani}}]{man83}%
  \BibitemOpen
  \bibfield  {author} {\bibinfo {author} {\bibfnamefont {E.}~\bibnamefont
  {Manousakis}}, \bibinfo {author} {\bibfnamefont {S.}~\bibnamefont {Fantoni}},
  \bibinfo {author} {\bibfnamefont {V.~R.}\ \bibnamefont {Pandharipande}}, \
  and\ \bibinfo {author} {\bibfnamefont {Q.~N.}\ \bibnamefont {Usmani}},\
  }\href {\doibase https://doi.org/10.1103/PhysRevB.28.3770} {\bibfield
  {journal} {\bibinfo  {journal} {Phys. Rev. B}\ }\textbf {\bibinfo {volume}
  {28}},\ \bibinfo {pages} {3770} (\bibinfo {year} {1983})}\BibitemShut
  {NoStop}%
\bibitem [{\citenamefont {Ceperley}(1995)}]{cep95}%
  \BibitemOpen
  \bibfield  {author} {\bibinfo {author} {\bibfnamefont {D.~M.}\ \bibnamefont
  {Ceperley}},\ }\href {\doibase 10.1103/revmodphys.67.279} {\bibfield
  {journal} {\bibinfo  {journal} {Rev. Mod. Phys.}\ }\textbf {\bibinfo {volume}
  {67}},\ \bibinfo {pages} {279} (\bibinfo {year} {1995})}\BibitemShut
  {NoStop}%
\bibitem [{\citenamefont {Rossoti}\ \emph {et~al.}(2017)\citenamefont
  {Rossoti}, \citenamefont {Teruzzi}, \citenamefont {Pini}, \citenamefont
  {Galli},\ and\ \citenamefont {Bertaina}}]{ros17}%
  \BibitemOpen
  \bibfield  {author} {\bibinfo {author} {\bibfnamefont {S.}~\bibnamefont
  {Rossoti}}, \bibinfo {author} {\bibfnamefont {M.}~\bibnamefont {Teruzzi}},
  \bibinfo {author} {\bibfnamefont {D.}~\bibnamefont {Pini}}, \bibinfo {author}
  {\bibfnamefont {D.~E.}\ \bibnamefont {Galli}}, \ and\ \bibinfo {author}
  {\bibfnamefont {G.}~\bibnamefont {Bertaina}},\ }\href {\doibase
  10.1103/PhysRevLett.119.215301} {\bibfield  {journal} {\bibinfo  {journal}
  {Phys. Rev. Lett.}\ }\textbf {\bibinfo {volume} {119}},\ \bibinfo {pages}
  {215301} (\bibinfo {year} {2017})}\BibitemShut {NoStop}%
\bibitem [{\citenamefont {Abolins}\ \emph {et~al.}(2018)\citenamefont
  {Abolins}, \citenamefont {Zillich},\ and\ \citenamefont {Whaley}}]{abo18}%
  \BibitemOpen
  \bibfield  {author} {\bibinfo {author} {\bibfnamefont {B.~P.}\ \bibnamefont
  {Abolins}}, \bibinfo {author} {\bibfnamefont {R.~E.}\ \bibnamefont
  {Zillich}}, \ and\ \bibinfo {author} {\bibfnamefont {K.~B.}\ \bibnamefont
  {Whaley}},\ }\href {\doibase 10.1063/1.5005522} {\bibfield  {journal}
  {\bibinfo  {journal} {J. Chem. Phys.}\ }\textbf {\bibinfo {volume} {148}},\
  \bibinfo {pages} {102338} (\bibinfo {year} {2018})}\BibitemShut {NoStop}%
\bibitem [{\citenamefont {Bertaina}\ \emph {et~al.}(2016)\citenamefont
  {Bertaina}, \citenamefont {Motta}, \citenamefont {Rossi}, \citenamefont
  {Vitali},\ and\ \citenamefont {Galli}}]{ber16}%
  \BibitemOpen
  \bibfield  {author} {\bibinfo {author} {\bibfnamefont {G.}~\bibnamefont
  {Bertaina}}, \bibinfo {author} {\bibfnamefont {M.}~\bibnamefont {Motta}},
  \bibinfo {author} {\bibfnamefont {M.}~\bibnamefont {Rossi}}, \bibinfo
  {author} {\bibfnamefont {E.}~\bibnamefont {Vitali}}, \ and\ \bibinfo {author}
  {\bibfnamefont {D.}~\bibnamefont {Galli}},\ }\href {\doibase
  10.1103/PhysRevLett.116.135302} {\bibfield  {journal} {\bibinfo  {journal}
  {Phys. Rev. Lett.}\ }\textbf {\bibinfo {volume} {116}},\ \bibinfo {pages}
  {135302} (\bibinfo {year} {2016})}\BibitemShut {NoStop}%
\bibitem [{\citenamefont {Cuervo}\ \emph {et~al.}(2005)\citenamefont {Cuervo},
  \citenamefont {Roy},\ and\ \citenamefont {Boninsegni}}]{cue05}%
  \BibitemOpen
  \bibfield  {author} {\bibinfo {author} {\bibfnamefont {J.~E.}\ \bibnamefont
  {Cuervo}}, \bibinfo {author} {\bibfnamefont {P.-N.}\ \bibnamefont {Roy}}, \
  and\ \bibinfo {author} {\bibfnamefont {M.}~\bibnamefont {Boninsegni}},\
  }\href {\doibase https://doi.org/10.1063/1.1872775} {\bibfield  {journal}
  {\bibinfo  {journal} {J. Chem. Phys.}\ }\textbf {\bibinfo {volume} {122}},\
  \bibinfo {pages} {114504} (\bibinfo {year} {2005})}\BibitemShut {NoStop}%
\bibitem [{\citenamefont {Rossi}\ \emph {et~al.}(2009)\citenamefont {Rossi},
  \citenamefont {Nava}, \citenamefont {Reatto},\ and\ \citenamefont
  {Galli}}]{ros09}%
  \BibitemOpen
  \bibfield  {author} {\bibinfo {author} {\bibfnamefont {M.}~\bibnamefont
  {Rossi}}, \bibinfo {author} {\bibfnamefont {M.}~\bibnamefont {Nava}},
  \bibinfo {author} {\bibfnamefont {L.}~\bibnamefont {Reatto}}, \ and\ \bibinfo
  {author} {\bibfnamefont {D.~E.}\ \bibnamefont {Galli}},\ }\href {\doibase
  10.1063/1.3247833} {\bibfield  {journal} {\bibinfo  {journal} {J. Chem.
  Phys.}\ }\textbf {\bibinfo {volume} {131}},\ \bibinfo {pages} {154108}
  (\bibinfo {year} {2009})}\BibitemShut {NoStop}%
\bibitem [{\citenamefont {Ceperley}(1996)}]{cep96}%
  \BibitemOpen
  \bibfield  {author} {\bibinfo {author} {\bibfnamefont {D.}~\bibnamefont
  {Ceperley}},\ }in\ \href@noop {} {\emph {\bibinfo {booktitle} {Monte Carlo
  and Molecular Dynamics of Condensed Matter Systems}}}\ (\bibinfo {year}
  {1996})\BibitemShut {NoStop}%
\bibitem [{\citenamefont {Aziz}\ \emph {et~al.}(1995)\citenamefont {Aziz},
  \citenamefont {Janzen},\ and\ \citenamefont {Moldover}}]{azi95}%
  \BibitemOpen
  \bibfield  {author} {\bibinfo {author} {\bibfnamefont {R.~A.}\ \bibnamefont
  {Aziz}}, \bibinfo {author} {\bibfnamefont {A.~R.}\ \bibnamefont {Janzen}}, \
  and\ \bibinfo {author} {\bibfnamefont {M.~R.}\ \bibnamefont {Moldover}},\
  }\href {\doibase 10.1103/physrevlett.74.1586} {\bibfield  {journal} {\bibinfo
   {journal} {Phys. Rev. Lett.}\ }\textbf {\bibinfo {volume} {74}},\ \bibinfo
  {pages} {1586} (\bibinfo {year} {1995})}\BibitemShut {NoStop}%
\bibitem [{\citenamefont {Schmidt}\ \emph {et~al.}(1981)\citenamefont
  {Schmidt}, \citenamefont {Lee}, \citenamefont {Kalos},\ and\ \citenamefont
  {Chester}}]{sch81}%
  \BibitemOpen
  \bibfield  {author} {\bibinfo {author} {\bibfnamefont {K.~E.}\ \bibnamefont
  {Schmidt}}, \bibinfo {author} {\bibfnamefont {M.~A.}\ \bibnamefont {Lee}},
  \bibinfo {author} {\bibfnamefont {M.~H.}\ \bibnamefont {Kalos}}, \ and\
  \bibinfo {author} {\bibfnamefont {G.~V.}\ \bibnamefont {Chester}},\ }\href
  {\doibase 10.1103/physrevlett.47.807} {\bibfield  {journal} {\bibinfo
  {journal} {Phys. Rev. Lett.}\ }\textbf {\bibinfo {volume} {47}},\ \bibinfo
  {pages} {807} (\bibinfo {year} {1981})}\BibitemShut {NoStop}%
\bibitem [{\citenamefont {Panoff}\ and\ \citenamefont {Carlson}(1989)}]{pan89}%
  \BibitemOpen
  \bibfield  {author} {\bibinfo {author} {\bibfnamefont {R.~M.}\ \bibnamefont
  {Panoff}}\ and\ \bibinfo {author} {\bibfnamefont {J.}~\bibnamefont
  {Carlson}},\ }\href {\doibase 10.1103/physrevlett.62.1130} {\bibfield
  {journal} {\bibinfo  {journal} {Phys. Rev. Lett.}\ }\textbf {\bibinfo
  {volume} {62}},\ \bibinfo {pages} {1130} (\bibinfo {year}
  {1989})}\BibitemShut {NoStop}%
\bibitem [{\citenamefont {Roberts}\ \emph {et~al.}(1964)\citenamefont
  {Roberts}, \citenamefont {Sherman},\ and\ \citenamefont {Sydoriak}}]{rob64}%
  \BibitemOpen
  \bibfield  {author} {\bibinfo {author} {\bibfnamefont {T.~R.}\ \bibnamefont
  {Roberts}}, \bibinfo {author} {\bibfnamefont {R.~H.}\ \bibnamefont
  {Sherman}}, \ and\ \bibinfo {author} {\bibfnamefont {S.~G.}\ \bibnamefont
  {Sydoriak}},\ }\href@noop {} {\bibfield  {journal} {\bibinfo  {journal} {J.
  Res. Natl. Bur. Stand.}\ }\textbf {\bibinfo {volume} {68A}},\ \bibinfo
  {pages} {567} (\bibinfo {year} {1964})}\BibitemShut {NoStop}%
\end{thebibliography}
\end{document}